# Towards Intelligent Context-Aware 6G Security

André N. Barreto[1], Stefan Köpsell[1], Arsenia Chorti[2], Bertram Poettering[3], Jens Jelitto[3], Julia Hesse[3], Jonathan Boole[3], Konrad Rieck[4], Marios Kountouris[5], Dave Singelee[6], Kumar Ashwinee[7]

[1] Barkhausen Institut, Dresden, Germany
[2] ETIS UMR 8051, CYU, ENSEA, CNRS, Paris, France
[3] IBM Research, Zurich, Switzerland
[4] TU Braunschweig, Germany
[5] Eurecom, Sophia Antipolis, France
[6] KU Leuven, Belgium
[7] Vrije Universiteit Brussel, Belgium

**Abstract:** Imagine interconnected objects with embedded artificial intelligence (AI), empowered to sense the environment, see it, hear it, touch it, interact with it, and move. As future networks of intelligent objects come to life, tremendous new challenges arise for security, but also new opportunities, allowing to address current, as well as future, pressing needs. In this paper we put forward a roadmap towards the realization of a new security paradigm that we articulate as *intelligent context-aware security*. The premise of this roadmap is that sensing and advanced AI will enable context awareness, which in turn can drive intelligent security mechanisms, such as adaptation and automation of security controls. This concept not only provides immediate answers to burning open questions, in particular with respect to non-functional requirements, such as energy or latency constraints, heterogeneity of radio frequency (RF) technologies and long life span of deployed devices, but also, more importantly, offers a viable answer to scalability by allowing such constraints to be met even in massive connectivity regimes. Furthermore, the proposed roadmap has to be designed ethically, by explicitly placing privacy concerns at its core. The path towards this vision and some of the challenges along the way are discussed in this contribution.

## 1. Introduction

We are entering a new era of interconnected intelligence of myriads of autonomous devices, equipped with enhanced sensing, processing, and learning capabilities. These networked intelligent systems will underpin the global functioning of our societies and enable formidable progress in the industrial, health, transportation, environmental and educational sectors. As emerging cyber-physical systems and the Internet of intelligent things (IoIT) will be key enablers for a plethora of tremendous services to society, by 2025, networks will have to support an estimated 60 billion sensor-assisted and often autonomous devices and vehicles (e.g., smartphones, robots, cars, drones, etc.). The security of **massive IoIT** systems, which will be highly **heterogeneous** (from RFIDs to critical infrastructure), will be manufactured by different vendors without homogeneous processes and will operate under unprecedentedly **diverse constraints** (power, latency, computational and memory resources), has to be addressed today. Two further challenges need to be taken into account; the extensive introduction of **artificial intelligence** (AI) and machine learning (ML), as well as the prospect of **quantum computing** expand both the breadth and the depth of the attack surface beyond anything witnessed insofar.

Confidentiality, integrity, accountability, access control and privacy must be at the core of today's designs of future generations of intelligent networks. In the past, context-agnostic security solutions were introduced as add-ons to network design choices. A break from this paradigm is needed for the design of future security controls, primarily because:

★ *Static security solutions cannot scale efficiently while meeting latency, computation, power constraints, heterogeneity and lack of homogeneous production procedures;*

★ *AI/ML introduces novel vulnerabilities, the full extent of which is yet unknown;*

★ *It seems feasible for quantum computing to become commercially available.*

The challenges ahead require a fundamentally different solution. To this end, looking beyond the current research horizon, a radically new, **context-aware, AI-empowered roadmap for the design of intelligent and adaptive security controls** is needed. This concept offers solutions to pressing needs, such as securing low-end IoT systems in the quality of security (QoSec) framework, or latency-constrained verticals, such as automotive and industrial IoT for securing future, autonomous, and heterogeneous hyperconnected systems under an augmented attack surface that accounts for quantum computing and AI. Only through this can we imagine a future in which neither users nor developers would need to worry about security and privacy; in which small innovative start-ups could concentrate on functionality and in which automation of security controls could alleviate the need for large security and legal departments, where users could enjoy new services without engaging with cumbersome and error-prone security and privacy settings in a longer-term vision of fully automated security.

Notably, having learned from the experience of the general data protection regulation (GDPR) in the European Union, an ethical and durable design cannot be neglected, as major legal, regulatory and ethical issues will arise, particularly with respect to the collection of huge amounts of sensing data. **Privacy** should be addressed by design; both by means of new engineering solutions (e.g., for sensing and localization) and through a new legal and regulatory framework, for instance, by giving people and institutions the right not to be sensed, are needed. Moreover, other important issues, like the liability for the decisions, the possibility of misuse and environmental impact must not be ignored.

However, this vision of intelligent context-aware security cannot be realized by incrementally adding new attributes to existing solutions. In addition to a new ethical framework, the community should work towards:

- *Security-oriented context awareness, e.g., by incorporating attention ML mechanisms and goal oriented risk assessment methodologies;*
- *Acknowledging the need for genuinely adaptive security controls at all network layers, stretching from post-quantum to physical layer and hardware security and context-aware device pairing;*
- *Interfacing network intelligence and security adaptivity in a vertical, security plane (orchestrator), offering the flexibility to allow even local security decision reasoning.*

This roadmap is aligned with ongoing discussions at the European Telecommunications Standardization Institute (ETSI) around autonomic security controls and aims at paving the way to reaching these ambitious targets. In our vision, multimodal fusion of sensed data with network descriptors will extract the security context, i.e., a picture of the threat level, requirements, and available measures. For each particular application and scenario, an orchestrator will dynamically select the best security measures from a wide palette of adaptive security controls and estimate the level of security that can be achieved.

## 2. Roadmap towards intelligent, context aware security

There are currently vivid discussions around the possibility of adjusting the security level according to the needs of different slices in a Quality-of-Security (QoSec) framework, to the concepts of trust and trustworthiness, to the impact of privacy issues, to the role of intelligence in future networks, to the potential exploitation of sensing in a wide array of applications, and so forth. However, existing solutions address only partially the breadth and the depth of the required intelligent security controls, without considering *lightweight post-quantum cryptography for critical infrastructure, provable and context-aware adaptive controls at all layers, measurable trustworthiness and hardening and privacy by design*. Similarly to a living organism, security has to respond autonomously to changing threat levels by exploiting a full body of security controls. This will require new science to implement the brain of the context-aware security, its senses, the novel security controls themselves, and the ethics that should govern it. In further detail, the following elements are pivotal

*Sensing* will be an integral part of future networks [1]. Besides the myriad of sensors already available in consumer devices, joint communication and sensing technologies, reusing frequency/time resources of communications systems will be pivotal in this. The availability of millimeter-waves (mmWave) raises the possibility of high-resolution 3D imaging and even spectroscopy. All this data can be fused to provide a complete map of the environment in multiple dimensions. Such aspects must be investigated along with reliability and trustworthiness of sensed data.

*Privacy* is a central concern in a world where diverse sensors continuously collect large amounts of fine-grained data. This affects not only users but also bystanders whose data might be collected implicitly, without outright consent. While techniques for privacy-preserving data analysis exist, research has shown that anonymizing high-dimensional data is difficult and can lead to privacy violations [2], especially under other constraints (e.g., low power). Doing it correctly while preserving data utility might not be achievable at all for high-dimensional datasets [3]. Hence, one challenge that needs to be addressed concerns finding the right balance between privacy and utility.

The concept of *context* serves as a base for context-aware applications, i.e., applications which can sense their environment and react accordingly. We consider context to be a construct that goes far beyond the pure description of the environment (e.g., in terms of where, when, what and who) and therefore context-awareness becomes closely related to situation-awareness, i.e., the process of perception of elements in the current situation, the comprehension of the current situation and the projection of future status [4]. State-of-the-art ML tools in natural language processing (NLP), control and human-machine interaction, i.e., attention models [5], goal-oriented analysis [6] and multimodal fusion [7] can be leveraged for context distillation. By jointly exploiting physical and cyber context awareness, we envision advanced intelligence in future networks, which, among other purposes, will serve security goals.

Looking further at **the use of ML**, particularly adversarial ML, we need to rethink properties such as security, privacy, robustness, explainability, and fairness. ML has been traditionally designed with static scenarios in mind, where data does not drift, and malicious entities do not try to undermine the system. Research in this direction is still in its infancy and while some of the properties can be achieved in limited scenarios, much work remains to be done to guarantee them in general conditions [8]. It also remains to be seen if all these desirable properties may even be realized mutually or if they are contradictory in some scenarios. Moreover, threat modelling is a critical piece of the puzzle when it comes to estimating risk and balancing exposure while applying ML to the security domain. Adversaries can be classified based on their background knowledge, whether they are active or passive and if they choose to attack during training or inference phase. We also need to consider whether the ML setting is centralized or decentralized.

*Adaptive security* is needed in the context of future wireless communication, where solutions are required to run on an extremely broad range of devices and in many different environments, unlike current solutions that are static, in the sense that they are optimized to solve a security challenge in one particular usage profile. Crucially, in future networks the profiles of devices may change over time. By adaptive security we understand the security that can be reached by a dynamic security engine that follows changing usage profiles and context, always aiming at delivering the best security possible in that moment. At this point in time, the adaptive security topic remains unexplored and any contributions will potentially open new research avenues.

*Quality of security (QoSec)* describes the "degree of security" in a holistic, measurable, and comparable way. Preliminary work in this direction was done in [9], later enriched in [10] and other more recent works on the joint optimization of QoSec and QoS [11]. We argue that a more holistic view on QoSec is needed, incorporating preventive security controls from different domains (e.g., physical layer security (PLS)), also including aspects of detection, reaction and recovery.

*Security building blocks* for future-proof security systems should incorporate quantum-resistant cryptographic algorithms. The NIST post-quantum competition addresses this issue and multiple round-3 finalists are currently being evaluated, such as [12]. However, these algorithms are relatively complex, and there is a clear need for novel post-quantum algorithms oriented towards low-cost embedded connected systems and for adaptive cryptographic algorithms and protocols that dynamically adjust their configuration and parameters according to the context. In realizing adaptive cryptosystems, one of the most challenging security problems is the initial trust establishment between devices. One approach is to leverage context during key establishment or device pairing. Several solutions in this domain have been proposed, for example based on physiological signals within a human body or on vibration within vehicles [13]. However, open research challenges include that some of these schemes are broken, while none of them can be applied in a large set of application scenarios. Finally, it is important that context-leveraged security is applied in an automatic way, i.e., without any manual interactions.

*Societal Impact* and the success of adaptive context-aware security measures depend not only on business creation, but also on providing a clear ethical, legal and societal (ELSA) framework, as well as on the engagement of local, national, and international authorities to enable adaptation of novel technologies into their constituencies. Based on current research, new technology will have to be developed in order to address all these issues. A high-level view of the science-to-technology conversion to reach the overall technological breakthrough of context-aware security is given in Table I.

**Table I: Science to technology conversion**

| New science | Conversion to new technology |
|---|---|
| **Sensing:** Joint communication & sensing at mmWave, sensor fusion, trustworthy sensing by design | Spectroscopy, 3D radar imaging, multimodal sensing, enhanced context awareness, privacy-preserving trust |
| **Context awareness:** Multimodal fusion of sensing and network data, network intelligence | Context extraction, knowledge plane, knowledge sharing between different domains, e.g., slices, operators |
| **Adversarial ML:** Security, privacy, robustness, | Estimating risk and balancing exposure while applying |

| explainability and fairness of ML | ML to the security domain |
|---|---|
| **Adaptive security:** Crypto with agile security level, controls at all layers, cross-layer security | Secure wireless links, peer-to-peer connections, devices organized as groups autonomously, e.g., platooning |
| **PLS and hardware security:** Codes with explicit reliability-equivocation, secret key generation, PUFs from context | Provable / quantifiable security, lightweight and low-latency hybrid PLS-crypto that adapts to PHY context |
| **Post quantum crypto:** New post quantum algorithms beyond those by project members | Post-quantum crypto for low-cost embedded connected systems, securing future critical infrastructure |
| **Trust building:** Initial trust between devices, use context during key establishment or device pairing | New device pairing and trust building, build automatic pairing mechanisms without manual interactions |
| **Privacy:** Trade-offs and balance in joint privacy and utility preservation for high-dimensional datasets | Ensure privacy, extending to bystanders who might be swept into the data collection without outright consent |
| **Societal impact:** Privacy by design, SEL protocol and risk and impact assessment | Ethical, legal and societal framework for context-aware security, engagement of local/ national/supranational authorities |

Novel adaptive and context-aware security controls should be introduced both from the domain of cryptography as well as that of physical-layer security (PLS) and hardware security. The mechanisms must not only allow reaching a set of static security levels expressed via QoSec, but must be adaptive by design (key sizes, key rates, etc.). Adaptivity should not only happen at the level of the algorithms and specific security mechanisms, as in existing works in QoSec, but also at higher levels, e.g., by substituting PLS by cryptographic schemes when the latter are advantageous, i.e., if the context requires this. Unlike in the past, privacy issues cannot be treated as standalone independent requirements, but must be integrated in the design. The ethical framework, jointly with the system architecture should form the context-aware security controls.

## 3. System architecture for quantifiable security

In this section we propose the architecture of a context-aware security solution and its main building blocks. One of the foundations is the notion of Quality of Security (QoSec), described in Section 2, which is a fundamental input to the closed-loop control of any context-aware security framework. QoSec is used on the one hand to specify the security requirements from an user and application perspective. On the other hand, QoSec is used to measure the actually provided degree of security, based on the context and the security controls in place. An orchestrator must adapt the security configuration, e.g. the set of adaptive security controls used as well as their configuration in order to adjust the current state (in terms of QoSec) to the desired target state. Thereby, the current context must be incorporated in the decision process. If the current context does not allow the target state to be reached, the user will be informed and measures from the domain of "graceful degradation" are taken. The proposed architecture is depicted in Fig. 1.

The orchestrator itself can be realized by different architectures, ranging from centralized to fully distributed orchestration. To derive actionable decisions from the context, we envision a multi-layer decision process for the orchestrator. At the higher layers more abstract security controls will be orchestrated to achieve the target QoSec (e.g. encryption for messages transmitted over a given public network link to achieve confidentiality), whereas at the lower layers the configuration of the selected abstract security controls will be adapted (e.g. choosing AES with 256 bit keys). The decision layers will span from conservative models, such as static policies and rules, to adaptive strategies, such as probabilities models and machine learning. The goal would be to estimate secure configurations in highly dynamic environments and at the same time provide fallback solutions in case of failures and attacks. For the static decision models, we propose to build on existing rule-based methods, while for the dynamic models, we propose to design novel decision concepts based on deep learning and attention models. Thereby this process could be distributed with different levels of granularity to support federated orchestration.

While dynamic orchestration of security modules could provide a novel level of protection for connected systems, it could also expose new attack surfaces to adversaries. To mitigate this risk from the start, a security-self assessment of the orchestrator is also needed. To this end, offensive security concepts would have to be directly integrated into the orchestrator and act as antagonists during its decision process. This approach could be explored with different concepts of offensive security, as well as strategies for realizing the antagonistic decision process.

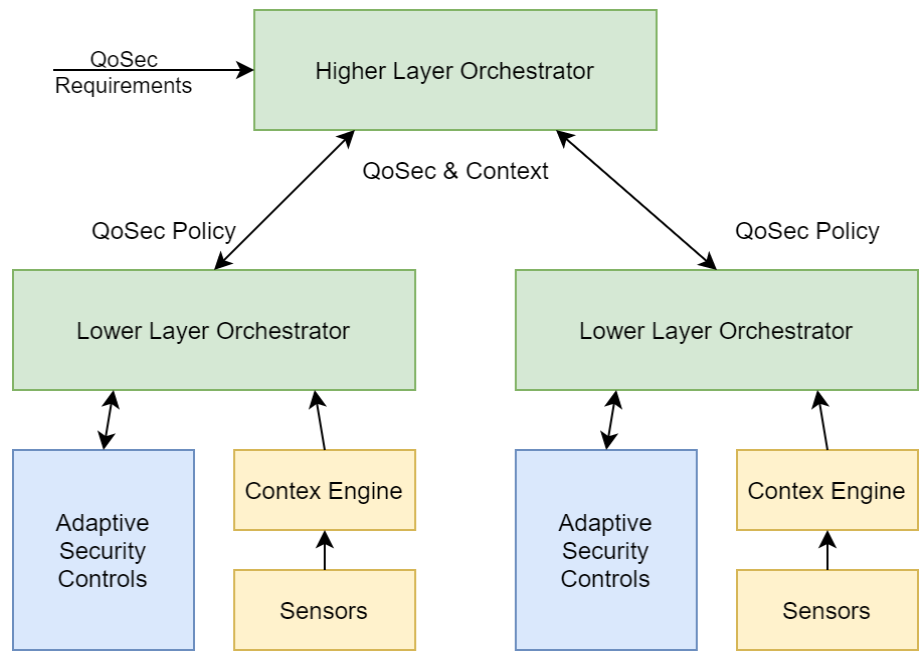

Figure 1: System Architecture

## 4. Sensing

Sensing integration will be one of the key enhancements in 6G [1]. It will enable an enhanced context awareness, which can be obtained through fusion of multiple sensor inputs. Cameras, radars, lidars, and (ultra)sound can be used to build a 3D dynamic map of the environment, or to identify objects, generate encryptionkeys, or even detect potential intruders. Radar range Doppler patterns and video can be used to classify security risks. Infrared sensors and microphones can detect heat and noise sources, which can be helpful in identifying attackers. Data from thermometers can identify if users are in the same environment. With sensor fusion, we can extract the physical context, but the use of such data for security applications has not been investigated yet in theliterature. We also have to consider the trade-off between cost, energy consumption and availability of sensors in relation to their usefulness for security-relevant context analysis.

Looking further ahead, integrated trustworthy RF sensing will be incorporated in intelligent networking. Atthe same time, the statistical characterization of the propagation channel provides information for security, particularly for PLS. Also, joint communications and radar sensing (JC&S) will be a key feature in 6G, with wireless communications and radar using the same waveforms and hardware, allowing 3D radar images. Together with improved positioning, this can be used to assess the security risks and the effectiveness of PLS [14]. It is also essential that the signals employed for positioning and RF sensing are robust against different active attacks, which could be achieved through a proper waveform and coding design.

The sensing information can be first used to extract low-level context, aligning with ongoing efforts on incorporating it in the form of *who, what, when, where¸* in the system intelligence. Central to future studies should be high precision indoor and outdoor localization with ML and statistical signal processing techniques. Leveraging the use of massive multiple input multiple output systems in sub-THz communications, we envision centimetre level indoor localization precision. Privacy preservation measures have to be taken, for instance with the use of state-of-the-art channel charting techniques.

Finally, sensing features and KPIs for each type of sensor for the security context have to be defined, with aspects such as precision, sensitivity, and resolution taken into consideration. Data from radar and cameras, that provide multi-dimensional information, must be compressed before being shared with the semantic fusion, as will be discussed in the following section. For PLS in particular, the properties of the propagation channel, such as the number of independent paths, the delay and angular spread, will be essential for estimating related metrics, e.g., the secrecy capacity and secret key rates. The interface towards each sensor has to be specified, i.e., which data should be provided and the required quality.

## 5. Reaching context awareness

Privacy-preserving context distillation will be a key requirement in future generations of wireless. With the proliferation of the number of devices, threats to privacy and user data will grow proportionately. Apart from primary contextual information with respect to location, node identity, time, etc., diverse input sources will in the future be fused sources to generate higher level contextual awareness. In this framework, privacy-preserving methods should be selected and implemented, based on the specificities of the context and focusing on privacy (e.g., categories / size of data, possibilities of identification, data flows, etc.). Furthermore, relevant principles such as privacy by design, ethics by design as well as data protection by design should be followed.

Overall, we should aim at distilling the context that is relevant to the security goals set out by the orchestrator. Leveraging attention mechanisms from NLP, we could jointly compress the sensing and cyber data by removing irrelevant to security goals information. Starting at the individual node level, we could learn node behavioral patterns, while at the network segment level capturing the dynamics of interactions between multiple nodes in order to characterize group processes. We should additionally incorporate ML tools from human-machine interaction and transformer models, while whenever appropriate, multi-modal fusion techniques could distil context jointly from sources of different modalities.

The key goal should be the development of a goal-oriented, context-aware security risk assessment, as a means to empower scalable, adaptive and efficient security in future hyperconnected intelligent systems. Leveraging certain aspects of the semantics of data, notably the importance and effectiveness of information for attaining the security-oriented goals will be instrumental here. In this context, semantics is related to both qualitative intrinsic (objective) and contextual, application-dependent (subjective) attributes of information. A starting point would be to provide a tractable and insightful mapping between information semantics and security functions, allowing us to design several, goal-driven, data representation, security violation detection and information prioritization mechanisms. A context-aware processing (for adaptively or partially securing heterogeneous data flows) and semantic risk assessment (by agilely orchestrating multi-source data collection and fusion) needs to be developed.

Finally, hardening by design must be a key aspect in order to render context distillation resilient by design against active attacks. Two parallel approaches can be taken in this direction. First, the sensing and data collection process itself must be studied from a security point of view, with the objective of deriving trustworthiness scores that express the level of resilience of the sensing process. Second, the ML techniques built should be analyzed and hardened. Different threats should be investigated, such as evasion, poisoning and inference attacks and robust defense mechanisms should be devised. The development must be conducted as an iterative process, where offensive and defensive tasks would be intertwined and jointly advance the overall security of ML algorithms in the project. The targeted context awareness should be hardened in its design, providing a solid foundation for adaptive security.

## 6. Adaptive security controls

In 6G ground, air and space communications will be integrated for the first time, and, therefore, there is now a wide consensus that security controls at all layers, including the PHY, should be developed for the 6G security protocols. There is a need for a toolbox of cryptographic and physical layer security (PLS) [15] solutions that can be incorporated based on the context and the QoSec requirements. In our proposed roadmap we have identified *quantum-secure communication, context-leveraged security, reactive cryptography* and *context-aware physical layer security* among the key building blocks for the development of adaptive security controls for future generations of wireless.

With respect to quantum-secure communications, new quantum-resilient protocols for basic cryptographic tasks, tailored to the requirements of future wireless devices should be designed. As an example, we envision hybrid key generation schemes that ensure information-theoretic secrecy provided that enough entropy is present in the channel, while guaranteeing computational secrecy in all cases. Such solutions could be tunable and thus deployable on devices with different capabilities. It should be further investigated whether quantum-computing capabilities can be used to better solve basic security tasks.

In what concerns context-leveraged security, future wireless technology allows for novel sensing capabilities that will let us bootstrap secure communication among edge devices without relying on, e.g., long-term secure storage or central trusted parties. Opportunistic pairing protocols for devices operated in similar context can be developed, by turning related context readings (movement, location, RF-sensing and optical inputs, etc.) into cryptographic keys. These protocols could secure peer-to-peer and group communication autonomously (platooning cars and drone swarms). Lightweight methods to leverage context to enhance trust, considering distributed consensus techniques to aid the decision process, could be developed.

In terms of reactive cryptography, future wireless systems will be highly dynamic, both in terms of device interconnections and in terms of resources (bandwidth, energy, etc.) available to them, which may change at high pace. This requires security solutions to be reactive, i.e., they must remain adaptive with respect to the services and guarantees that they deliver the required KPIs. Current cryptographic building blocks do not meet such requirements. We envision a refinement of established crypto notions, elevating context adaptivity to a prime goal. Provably secure solutions should be designed for multiple types of crypto primitives, most prominently adaptive authentication and key establishment, that are prepared to tolerate constantly changing participant sets and adjust the targeted security strength such that resources are optimally used at all times.

Finally, physical layer security is gaining a lot of attention in recent years. We should leverage the physical layer of future wireless devices for context-aware security, e.g. by extracting physical unclonable functions (PUFs) from intrinsic and extrinsic sources. Furthermore, decentralized identifiers (DIDs) and keys could be derived in different scenarios depending on the channel statistical properties, incorporating context awareness. At the same time, extrinsic PUFs could be used to expand context-aware security. Furthermore, randomness extraction from continuous random sources using fuzzy (lattice-based) extractors as well as short block length wiretap codes should be investigated, benchmarking them against information-theoretic second-order secrecy rate bounds. PLS performance is however strongly dependent on the channel properties and attacker models, and knowledge about these can be obtained as part of the context. Finally, of special interest could be novel hybrid crypto-PLS security controls, incorporating RF fingerprinting as an early (soft) authentication.

## 7. Societal, Ethical and Legal Requirements

This holistic approach towards an intelligent context-aware security architecture must include a rigorous privacy framework, or else it will not be embraced by society. This can be built primarily on three pillars:

i) *A fundamental rights framework*: the applicable framework regarding the relevant fundamental rights (with special attention to the right to privacy and the right to the protection of personal data as well as gender equity), should serve as a foundation for the system design;

ii) *An ethical and societal framework*: we should identify and analyse the applicable ethical principles rooted in the (regional) cultural heritage, including risks such as stigmatization and discrimination that may arise from any practices;

iii) *Regulatory frameworks relevant to the security application:* the security solution will have to rely on several information sources throughout its operation. The relevance of legal frameworks must then be thoroughly mapped, with special attention to data protection and security frameworks (e.g., such as these at the EU), and selected national frameworks. It is also necessary to look where liability may rest in the event of problems arising. The technology must be developed from the beginning with these three pillars in mind, satisfying their requirements and providing adequate information for all their needs.

Finally, a tailored-down impact assessment must be performed in order to generate information useful for all stakeholders, assisting them in taking informed decisions regarding the development of novel technological solutions. In particular, assistance is needed with respect to identifying and recommending the implementation of various safeguards and changes in the system architecture (e.g., privacy by design, ethics by design, etc.), ensuring adherence to a wide range of non-technical requirements. These guidelines must address legally compliant and ethically sound development, testing and use of future systems. They will focus inter alia on ethics, privacy preservation, data protection and legal compliance protocols.

## 8. Conclusions

In this paper we present a roadmap for reaching context-aware, adaptive security. Our proposal builds on major enhancements envisioned for beyond 5G systems, in particular with respect to advanced AI capabilities and sensing, that can play a key role in reaching context awareness and driving adaptivity of security controls. Importantly, we identify ethically responsible uses of sensing and AI and propose trustworthy sensing technologies to harvest context. The harvested environmental context can be fused with application and network descriptors to reach trustworthy context awareness with robustness to adversarial machine learning. Subsequently, a rich palette of adaptive security controls is proposed, ranging from novel post-quantum and auto-adaptive cryptographic algorithms to context-aware cutting-edge device pairing, hardware and physical layer security. Context awareness and adaptive security controls are finally interfaced to reach the desired autonomic adaptivity for intelligent context-aware security.